\documentstyle[12pt]{article}
\input epsf
\newcommand{\be}{\begin{equation}}
\newcommand{\ee}{\end{equation}}
\newcommand{\bea}{\begin{eqnarray}}
\newcommand{\eea}{\end{eqnarray}}
\newcommand{\lb}{\label}
\newcommand{\I}{\mbox{i}}
\newcommand{\D}{\mbox{d}}
\newcommand{\E}{\mbox{e}}

\begin{document}
\begin{titlepage}
\begin{flushright}
Freiburg THEP-98/7\\
gr-qc/9805014
\end{flushright}
\begin{center}
{\large\bf  EMERGENCE OF CLASSICALITY FOR PRIMORDIAL FLUCTUATIONS:
            CONCEPTS AND ANALOGIES}
\vskip 1cm
{\bf Claus Kiefer}
\vskip 0.4cm
 Fakult\"at f\"ur Physik, Universit\"at Freiburg,\\
  Hermann-Herder-Stra\ss e 3, D-79104 Freiburg, Germany.

\vskip 0.7cm
{\bf David Polarski}
\vskip 0.4cm
 Lab. de Math\'ematiques et Physique Th\'eorique, UPRES A 6083 CNRS\\
  Universit\'e de Tours, Parc de Grandmont, F-37200 Tours, France.\\
\vskip 0.3cm
 D\'epartement d'Astrophysique Relativiste et de Cosmologie,\\
 Observatoire de Paris-Meudon, F-92195 Meudon Cedex, France.

\end{center}
\date{\today}
\vskip 2cm
\begin{center}
{\bf Abstract}
\end{center}
\begin{quote}

We clarify the way in which cosmological perturbations of quantum origin, 
produced during inflation, assume classical properties. 
Two features play an important role in this process:
First, the dynamics of fluctuations which are presently on large cosmological 
scales leads to a very peculiar state (highly squeezed) that is
indistinguishable, in a precise sense, from a classical stochastic process.
This holds for almost all initial quantum states. 
Second, the process of decoherence by interaction with 
the environment distinguishes the field amplitude basis as the robust pointer 
basis. We discuss in detail the interplay between these features and use 
simple analogies such as the free quantum particle to illustrate the main
conceptual issues.

\end{quote}

\end{titlepage}

\section{Introduction}
The combination of particle physics models with general relativity
provides one with the possibility to construct quantitative
scenarios of the very early Universe. 

One important scenario, which helps to solve some of the outstanding 
problems of standard Big Bang cosmology like the homogeneity and the flatness 
problems, 
is that the Universe went through a stage of accelerated expansion,
an ``inflationary stage", in the very early part of its evolution \cite{lin}.
The inflationary scenario not only looses the dependence on peculiar
initial conditions, it also provides a quantitative way to understand
the formation of structure (galaxies and clusters of galaxies).
Indeed, in this scenario, the
origin of these large-scale structures can be traced back to vacuum 
{\em quantum} fluctuations of scalar fields \cite{haw82} and the resulting 
scalar
(gravitational potential) fluctuations of the metric. These fluctuations 
then lead eventually to the formation of large-scale structure in the 
universe and leave also their imprint as anisotropies in the cosmic background 
radiation. The anisotropies on large angular scales ($l\leq 20$, where
 $l$ is the multipole number) were 
detected by the COBE satellite.
Future satellite missions, Planck Surveyor \cite{pl} and MAP \cite{ma}, are 
scheduled for detection and high precision measurement of these anisotropies 
up to small angular scales (large $l$'s) and will enable us to possibly test 
the above scenario. 
(A recent critical review of some of these aspects is \cite{BG}.)
In addition inflation makes the important prediction of a background of 
relict gravitational waves \cite{al79} originating from tensor quantum 
fluctuations of the metric -- this constitutes an effect of linear quantum 
gravity.

Though research in this field has entered an exciting stage in which 
concrete models can be confronted with observations of ever increasing 
accuracy, an important question of principle is whether and to what extent 
the quantum origin of the primordial fluctuations can be recognised in the 
observations. This can only be answered after a thorough understanding of the
quantum-to-classical transition for the primordial fluctuations
has been achieved. Moreover, such an analysis is anyway necessary 
for an investigation into the possibilities to observe genuine
quantum gravitational effects {\em beyond} the linear approximation.
Such effects may arise, for example, from quantum gravitational
correction terms to the functional Schr\"odinger equation
\cite{KS} and may in principle be observable in the spectrum of the 
microwave background \cite{Ro}.

As a result of the dynamics of the fluctuations produced during 
inflation, one obtains for almost all initial quantum states a quantum state 
that is both highly squeezed and highly WKB \cite{AA94}.  
The peculiarity of this highly WKB state is characterised 
in the Heisenberg picture by the fact that 
the information about the initial momentum becomes lost -- a direct 
consequence of the vanishing of the decaying mode. This was shown for an 
initial vacuum (Gaussian) state \cite{PS1} as well as for initial number 
eigenstates \cite{PS2}. As a result, the fluctuations cannot be 
distinguished from a classical stochastic process, up to a tremendous 
accuracy well beyond the observational capabilities.  
This property does not require any environment \cite{PS1,PS2}.

However, interaction with the environment is unavoidable.
This was already stressed regarding
the problem of the entropy of the fluctuations
\cite{PS1,LPS}. Usually, classical properties for a certain system emerge by 
interaction of this system with its natural environment, a process referred 
to as 
decoherence (see \cite{dec} for a comprehensive review). 
It is therefore important to investigate its importance in the early 
Universe, since the fluctuations of the scalar field and the metric 
will most likely interact with various other fields.
Highly squeezed states are extremely sensitive to 
even small couplings with other fields (see Sect.~3.3.3 in \cite{dec}).
Since almost all realistic couplings are in field space (as opposed to the
field momentum), the distinguished ``pointer variable" (defining
classicality) is the {\em field amplitude} which also defines
a quantum nondemolition variable in the high squeezing limit \cite{KPS}. 
Interferences between different field amplitudes are therefore suppressed in 
the system itself with the same precision with which the non-diagonal 
elements of the density matrix describing the system can be taken zero. 
Environment-induced decoherence is effective when this precision is well 
beyond observational capabilities.

In \cite{KPS} we have stressed that these two features
play a decisive role in the emergence of classicality.
In the present article we shall give more
quantitative details than in the above mentioned ones about the nature 
of this quantum-to-classical transition. We shall present at length 
some aspects of the free quantum particle which, surprisingly, exhibits 
many features analogous to primordial fluctuations and discuss some 
physical ``experiments''. Our paper is organised as follows. Section~2 gives 
a brief
review of the dynamics of cosmological perturbations and clarifies
the first of the above two ingredients in the quantum-to-classical transition.
Section~3 then explains in what precise sense the system is indistinguishable 
from a classical stochastic process; this takes place up to an 
accuracy not only well beyond observational capabilities, but even well 
beyond the level of accuracy which is meaningful (beyond this accuracy, many 
other corrections should anyway be taken into account, as stressed
 in ~\cite{PS1}).
In Section~4 we present the analogy with the free quantum particle.
We discuss both the similarities to and differences from the
case of primordial fluctuations.
Section~5 gives a detailed account of how environment induced decoherence 
works for the primordial fluctuations. In particular, the rate of
de-separation as a measure for quantum entanglement is calculated for 
various initial states.
Section~6 gives our conclusions.

\vskip 4mm

\section{Dynamics of cosmological perturbations}

We start by giving a brief overview of primordial 
quantum fluctuations and the arguments put forward in \cite{PS1}.
Some of the explicit expressions will be needed for the calculation of the 
rate of de-separation in Sect.~5.
The simplest example, which nevertheless contains the essential features of 
linear cosmological perturbations, is that of a real massless (minimally 
coupled) scalar field $\phi$ in a
 ${\cal K}=0$ Friedmann Universe with time-dependent
scale factor $a$, whose action $S$ is given by 
\be
S=\frac{1}{2} \int \sqrt{-g} ~\partial^{\mu}\phi ~\partial_{\mu}\phi~.
\ee
It is crucial that during inflation one has an 
accelerated expansion. While the case considered here can be readily applied 
to gravitational waves (GW) or tensorial perturbations, all results can be 
extended to the case of scalar perturbations of the metric as well \cite{PS2}.
It turns out convenient to introduce the rescaled variable $y\equiv a\phi$
(the corresponding momenta thus being related by $p_y=a^{-1}p_{\phi}$; in 
the following we shall just write $p$ for $p_y$).
As a result of the coupling with the 
gravitational field, which plays here the role of an external classical 
field, we get squeezed quantum states \cite{Gr}. The dramatic consequences 
are best 
seen for a state $|0,~\eta \rangle$ which is the vacuum state of the field at 
some given initial time $\eta =\eta_0$ near the onset of inflation:
It will no longer be 
the vacuum at later time. Indeed, field quanta can be produced in pairs with 
opposite momenta, and one gets a {\em two-mode squeezed state}.

When the system can be thought of as being enclosed in a finite volume, the 
action $S$ becomes after Fourier transformation
\be
S = \sum_{\bf k} S_{\bf k}~,
\ee
where $S_{\bf k}$ is the action 
for the Fourier transform $y_{\bf k}$ corresponding to a given wave number 
${\bf k}$ and satisfying the Klein-Gordon equation  
\begin{equation}
y_{\bf k}'' +\left(k^2-\frac{a''}{a}\right)~y_{\bf k} = 0~.\label{KG}
\end{equation}
The prime denotes the derivative with respect to conformal time
$\eta\equiv \int  dt/a$.
The Hamiltonian can be similarly decomposed as a sum
of Hamiltonians for each mode,
\be H=\sum_{\bf k} H_{\bf k}~, \ee
where
\be H_{\bf k}= \frac{1}{2}\left(p_{\bf k} p_{\bf k}^* + 
k^2y_{\bf k}y_{\bf k}^* +
    \frac{a'}{a}\left[y_{\bf k}p_{\bf k}^* + p_{\bf k}y_{\bf k}^*
\right]\right) \ , \lb{ham} \ee
and $p_{\bf k}$ is given by 
$p_{\bf k}=y_{\bf k}'-\frac{a'}{a} y_{\bf k}$; it is the Fourier transform 
of $p=y'-\frac{a'}{a} y$, the momentum conjugate to $y$.
The field $y({\bf x})$ is real, hence, though $y_{\bf k}$ is complex,
 it satisfies 
$y_{-\bf k}=y_{\bf k}^*$, and the same applies to the Fourier transform of 
any real quantity. 
The Hamiltonian (\ref{ham}) can be formally viewed as the Hamiltonian of a
``time-dependent", possibly inverted, harmonic oscillator, 
the time-dependence coming from the changing scale factor. This fact makes 
the discussion formally very similar to many problems in quantum optics
\cite{Gr,Sch}. In fact, the $a$-dependent term in (\ref{ham}) leads to a
two-mode squeezed state, the modes being ${\bf k}$ and
$-{\bf k}$. We shall only consider states which are invariant under the 
reflection ${\bf k}\to -{\bf k}$. As a result, the sum needs to be taken over 
half of Fourier space also at the quantum level.

We want to look first at the system in the {\em Heisenberg 
representation}.
We have, introducing the field modes $f_k(\eta)$ with $f_k(\eta_0)=
1/\sqrt{2k}$,
\begin{eqnarray}
y_{\bf k}(\eta)
&\equiv& f_k(\eta)~a_{\bf k}(\eta_0)+f_k^*(\eta)~a^{\dag}_{-{\bf k}}(\eta_0)
\nonumber\\
&=&\sqrt{2k}~f_{k1}(\eta)y_{\bf k}(\eta_0)-\sqrt{{2\over k}}f_{k2}(\eta)
p_{\bf k}(\eta_0)~,\label{yk}
\end{eqnarray}
where $f_{k1}$ and $f_{k2}$ are the real and imaginary part of
$f_k$, respectively, and $a_{\bf k}$, $a_{\bf k}^{\dagger}$ denote
the standard annihilation and creation operators. 
Introducing further the momentum modes $g_k(\eta)$
 with $g_k(\eta_0)=\sqrt{{k\over 2}}$, we have
\begin{eqnarray}
p_{\bf k}(\eta)
&\equiv& -\I~\bigl\lbrack g_k(\eta)~a_{\bf k}(\eta_0)
-g_k^*(\eta)~a^{\dag}_{-{\bf k}}(\eta_0)\bigr\rbrack\nonumber\\
&=&\sqrt{{2\over k}}~g_{k1}(\eta)p_{\bf k}(\eta_0)+ 
\sqrt{2k}~g_{k2}(\eta)y_{\bf k}(\eta_0)~,\label{pk}
\end{eqnarray}
where $g_k\equiv g_{k1}+\I g_{k2}$.
The field modes $f_k$ obey (\ref{KG}) and are further constrained to 
satisfy the condition
\begin{equation}
2 ( f_{k1}g_{k1} + f_{k2}g_{k2} ) = 1~,\label{Wron}
\end{equation}
which can be viewed as either the Wronskian condition for (\ref{KG}) or the 
commutation relations imposed by canonical quantisation.
The relations (\ref{yk}) and (\ref{pk}) define, of course,
a Bogolubov transformation.
It should be stressed that all the equations (\ref{yk},\ref{pk},\ref{Wron}) 
are valid for the corresponding classical system, too, with the obvious 
difference that the quantities $y_{\bf k},~p_{\bf k},~a_{\bf k}$ are no 
longer operators.
Special interest will be attached to the quantity 
$B(k)\equiv f^*_k g_k$, whose real and imaginary parts read 
\begin{eqnarray}
\Re B(k) &=& \frac{1}{2}\ , \\
\Im B(k)=f_{k1}g_{k2}-f_{k2}g_{k1} &\equiv& F(k)\ . \lb{B}
\end{eqnarray}
The evolution of the system is conveniently parametrised by the squeezing 
parameter $r_k$, the squeezing angle $\varphi_k$ and the
rotation phase $\theta_k$.
The squeezing parameter $r_k$ can always be taken positive.
In terms of these parameters, the field modes can be cast in the form
\begin{eqnarray}
\sqrt{2k} f_k &=& {\rm e}^{-i(\theta_k+\varphi_k)}~(\cos \varphi_k
\E^{r_k}+\I\sin\varphi_k\E^{-r_k})
\label{f}\\
\sqrt{\frac{2}{k}} g_k &=& \I{\rm e}^{-i(\theta_k+\varphi_k)}~
(\sin \varphi_k~{\rm e}^{r_k}-\I\cos \varphi_k~{\rm e}^{-r_k})\ ,~~~~~~r_k>0.
 \lb {squeeze} \end{eqnarray}
The crucial point is that for large squeezing, $|r_k|\gg 1$, the
rotation phase and the
squeezing angle of the modes do not evolve independently of each other 
\cite{PS1}, but
allow to impose the condition
\begin{equation}
\theta_k+\varphi_k \simeq 0\ .\label{phase}
\end{equation}
The field modes $f_k$ ($g_k$) can be made real 
(imaginary) in the limit $|r_k|\to \infty$ with a {\it time-independent} 
phase rotation
\begin{eqnarray}
f_k &\to& e^{i\delta}~f_k\ , \nonumber\\
g_k &\to& e^{i\delta}~g_k~.\label{rot}
\end{eqnarray}
We can take $f_{k1}$ asymptotically positive without loss of generality.
Note that the quantity $B(k)$ is invariant under (\ref{rot}).
This corresponds to the following important property for the solutions 
$f_k$ and $g_k$ : Modes $f_k$ that leave the horizon
during the inflationary expansion and become bigger than the Hubble radius 
(horizon) can be written as a sum of a decaying mode ($\propto f_{k2}$) and 
a growing (``quasi-isotropic") mode ($\propto f_{k1}$) \cite{PS1}. 
During inflation, the decaying
mode becomes vanishingly small. Therefore one can make $f_k$ ($g_k$)
real (purely imaginary) in this limit. One then recognises from
(\ref{yk}) and (\ref{pk}) that $y_{\bf k}$ and $p_{\bf k}$
{\em commute} in this limit:
\be [y_{\bf k}(\eta), p_{\bf k}(\eta)] \approx 0\ . \lb{com} \ee
Moreover, it also means that $y_{\bf k}(\eta)$ commutes at
different times,
\be [y_{\bf k}(\eta_1), y_{\bf k}(\eta_2)] \approx 0\ . \lb{QND} \ee
This crucial property is transparent in the Heisenberg representation.
It has been emphasised in \cite{KPS} that this is the condition
for a quantum nondemolition measurement, a concept that is well
known in quantum optics (see, e.g., \cite{WM}). Observables obeying
(\ref{QND}) allow repeated measurements with great predictability, and we 
shall return to this point in Section~5.

In the {\em Schr\"odinger representation}, the 
dynamical evolution is given by the Schr\"odinger equation
\be \I\psi_k'(y,\eta)=\hat{H}_k\psi_k(y,\eta) \ . \ee
Let us consider the initial state to be the vacuum state $|0 \rangle$ 
at some initial time $\eta_0$ \cite{KPS}.
During the evolution, this initial Gaussian state stays a
Gaussian, but becomes {\em highly squeezed} due to the expansion
of the scale factor for modes bigger than the Hubble radius.
The squeezing is the Schr\"odinger analogue of the Bogolubov
transformation in the Heisenberg picture. 
The wave function in the amplitude (position) representation for given 
${\bf k}$ can be written as 
\be 
\Psi_{{\bf k}0}= \psi_{{\bf k}0}(y_{{\bf k}1}, \eta)~\psi_{{\bf k}0}
(y_{{\bf k}2}, \eta)~
\ee
with
\be 
\psi_{{\bf k}0}(y_{{\bf k}\alpha}, \eta)=\left(\frac{2\Omega_R}{\pi}
 \right)^{1/4}
      \exp\left(-\Omega(\eta)y_{{\bf k}\alpha}^2\right)
      \label{psi}~~~~~~~\alpha=1,2\; ,
\ee
where $\Omega_R\equiv \Re\Omega$, and we have further adopted the notation 
$\Re y_{\bf k}\equiv y_{{\bf k}1},~\Im y_{\bf k}\equiv y_{{\bf k}2}$. 
\par
The wave function (\ref{psi}) can be written in the following form,
where the Heisenberg mode $f_k(\eta)$ from (\ref{yk}) appears explicitly 
in the width of the Gaussian 
\cite{PS1},
\be
\Psi_{{\bf k}0}=\left( \frac{1}{\pi~|f_k|^2}\right)^{1/2}  
      \exp\left(-\frac{1-\I 2F(k)}{2~|f_k|^2} |y_{\bf k}|^2\right)\label{Psi}  
\ee
with the obvious identification $\Omega_R=\frac{1}{2}|f_k|^{-2}$ and
$\Omega_I\equiv\Im\Omega=-F(k)~|f_k|^{-2}$
(with $F(k)$ from (\ref{B})). The presence of the
growing mode in (\ref{yk}) thus directly leads to the broadening
of the wave function in the position representation.
It is convenient to introduce again the squeezing parameters
$r$ and $\varphi$ for
the state (\ref{Psi}). One has (cf. (\ref{B}) and
 (\ref{f}))\footnote{Note that, in contrast to standard conventions,
$\varphi=0$ corresponds here to squeezing in momentum,
not in position.}
\bea \vert f_k\vert^2 &=&\frac{1}{2k}\left(\cosh 2r_k+\cos 2\varphi_k
      \sinh 2r_k\right)\ , \lb{sq1} \\
      F(k) &=& \frac{1}{2}\sin 2\varphi_k\sinh 2r_k \ . \lb{sq2} \eea

The limit of large squeezing, which is obtained during
inflation, corresponds to $|F|\gg 1$.
We note that the state (\ref{Psi}) is annihilated by the following 
time-dependent operator \cite{PS1}
\be A_{\bf k}(\eta)=\frac{1}{\sqrt{2}}\Omega_R^{-1/2}(\Omega y_{\bf k}
    +\I p_{\bf k})\ . \ee
This plays the role of the annihilation operator for time-dependent
Gaussian states \cite{Ja}. For $\eta\to\eta_0$, $A_{\bf k}(\eta)$ becomes 
of course the annihilation operator $a_{\bf k}(\eta_0)$ in (\ref{yk})
and (\ref{pk}).

\par
The classical action $ S_{{\bf k},cl}$ i.e. the action $ S_{\bf k}$ 
introduced above and evaluated along the classical trajectory, is given by 
\be
 S_{{\bf k},cl}=y_{\bf k}~p^*_{\bf k}|^{\eta_1}_{\eta_2}~.
\ee
It is {\it not} the naive action that one would get for an oscillator with 
time-dependent frequency $k^2-\frac{a''}{a}$, but rather differs from it by a 
boundary term; this is hidden in the definition of $p_{\bf k}$.

Consider now instead of (\ref{Psi}) an arbitrary initial state
 at time $\eta_0$ of the form
\be
\vert\Psi\rangle=\sum_N c_N~|N \rangle \ , \label{sum}
\ee
where $|N \rangle$ denotes a state that initially,
 at a time $\eta_0$ when the modes are inside the 
Hubble radius, contains
 $N$ particles for each momentum $\bf k$ and $-{\bf k}$. 
In the amplitude (position) Schr\"odinger representation, $| N\rangle$ has 
the following expression \cite{PS2},
\be
\Psi_{{\bf k}N}=\left( -\frac{f_k}{f_k^*}\right)^N~
L_N\left( \frac{|y_{\bf k}|^2}{|f_k|^2}\right)~\Psi_{{\bf k}0}~,
\ee
where $L_N$ denotes a Laguerre polynomial.
This, in turn, implies the following form for $\Psi_{{\bf k}N}$ in the limit 
$|r_k|\to \infty$:
\be
\Psi_{{\bf k}N}\to R_N~\exp\left({\frac{\I}{\hbar}~S_{{\bf k},cl}}\right)~,
\label{wkb}
\ee
where $R_N$ is some real function. Note that we have used here the crucial 
property stated in (\ref{rot}). 
Therefore, in this limit an {\it almost arbitrary} initial state of the form 
(\ref{sum}) will enter the WKB regime. As in physical applications $|r_k|$ is 
of course not infinite, in the case of inflation ``almost arbitrary'' means 
that we exclude states with very special initial conditions, for example 
states that are initially (at the onset of inflation) extraordinarily squeezed 
in $y$ in such a way, that they exhibit no squeezing at all at late 
times.\footnote{Note the analogy of the high squeezing with Arnold's
cat map in classical mechanics.}
Inflation is itself based on ``natural" initial conditions, the
so-called no hair-conjecture, which prevents sub-Planckian modes
to yield a significant influence (see Chap.~9 in \cite{BG}). 
Analogously, one could call the exclusion of the just mentioned states a
``quantum no hair-conjecture", though one has to remember that such states are 
anyway {\it not} self-consistent with most inflationary models. 

\section{Equivalence with a classical stochastic process}

We shall now explain in several detailed ways in what operational sense our 
system cannot be distinguished from a classical stochastic process. 
In the Heisenberg representation, the physical arguments are transparent.
As a result of the dynamics and the resulting squeezing there is an 
almost perfect correlation between $y_{\bf k}$ and $p_{\bf k}$: For each 
given ``realisation" $y_{\bf k}$, we  
have $p_{\bf k}\simeq \frac{g_{k2}}{f_{k1}}~y_{\bf k}\equiv p_{{\bf k},cl}$, 
the {\it classical} momentum for $|r_k|\to \infty$. 
This can also be seen 
in the Schr\"odinger representation with the help of the Wigner function $W$. 
In the case of the above
Gaussian state (\ref{Psi}), one finds for the Wigner function
(often dropping $k$ in the following for simplicity),
\be 
W(y_1,{\bar p}_1,\eta)= \frac{1}{\pi}\exp\left(-\frac{y_1^2}
    {\vert f\vert^2}\right)\ \exp\left(-\vert f\vert^2~
     [{\bar p}_1-\frac{2F}{\vert f\vert^{2}}y_1]^2\right)~, \lb{W1} 
\ee 
and an analogous expression for $W(y_2,{\bar p}_2,\eta)$.
The quantity ${\bar p}_1~,{\rm resp.}~{\bar p}_2$, is canonically conjugate
to $y_1~,{\rm resp.}~y_2$, and so ${\bar p}_1=2~p_1,~{\bar p}_2=2~p_2$.
For large but finite squeezing, the Wigner function tells one that 
concrete values of the phase space variables
 are typically found inside an elongated ellipse 
(actually two identical ellipses for the real and imaginary parts of the 
canonical variables) in phase space 
with semimajor axis $\Delta y$ and semiminor axis $\Delta P$, 
where $P\equiv p-p_{cl}$.
In the limit $|r_k|\to \infty$ one has
\be 
W\to |\Psi_{\bf k}|^2~\frac{1}{4}~\delta
    \left(\frac{F}{|f|^2}y_{1}-
      p_{1}\right)~
\delta \left(\frac{F}{|f|^2}y_{2}-p_{2}\right)\ , \lb{W2}
\ee
where the delta distributions are defined with respect to $p_{1}$ 
and $p_{2}$, respectively.
This last result can be extended to more general initial states \cite{PS2}.

Let us consider now the concrete case of a quasi-de~Sitter expansion 
(${\dot H}\ll 3H^2$). In this case the modes are well-known and the following 
results are obtained in the long wavelength regime, i.e. for wavelengths much 
larger than the Hubble radius,
\bea
\Delta y &=& C_1~a\ , \\
\Delta P &=& \frac{|C_2|}{a}\ ,
\eea 
where 
\be
C_1=\frac{H_k}{\sqrt{2k^3}}\ , ~~~~~~~~
        C_2=-\I\frac{k^{3/2}}{\sqrt{2}H_k}~.\label{C}
\ee
In (\ref{C}), $H_k$ is the Hubble parameter evaluated at time $t_k$ with 
$k=a(t_k)H_k$, i.e. when the perturbation ``crosses'' the Hubble radius.
We note the relation 
\be
C_1 \Im C_2=-1/2\label{coh}
\ee
which follows from the commutator relations and which involves {\it both} 
modes. We see also that the typical volume in phase space remains constant 
\cite{LPS}. For a perturbation which will now appear on large cosmological 
scales, extreme squeezing is obtained, resulting in a ratio between the 
amplitude of decaying resp. growing mode, 
$f_{k2}$ resp. $f_{k1}$, that is proportional to $\exp(-2r_k)$. This is of
 the order $10^{-100}$ or less for the largest cosmological scales! 
Note that both real and imaginary parts of the field modes oscillate, 
hence one should not call them anymore growing and decaying modes. 
Still, they are very different because of a huge difference in amplitude. 
This is precisely due to the almost complete disappearance of the 
decaying mode, a result of the dynamics of the modes 
when they are outside the Hubble radius, i.e. when their wavelength is 
larger than the Hubble radius. Calling $A_k$ the ratio of these amplitudes, we 
have 
\be
A_k = \alpha_k~s_k^2\ ,
\ee
where $s_k$ is the relative stretching of the mode during its journey outside 
the Hubble radius, while the parameter $\alpha_k$ depends on the evolution of 
the background at both horizon (Hubble radius) crossings.
This shows that squeezing is operating as long as the modes are outside the 
Hubble radius, not necessarily during the inflationary phase.
Note also that the imaginary part can play a significant role in some cases, 
preventing the presence of zeroes in the fluctuations power spectrum 
\cite{PLB95}. This means that a power spectrum which has zeroes {\it cannot} 
be of quantum mechanical origin.
\par
Let us see now when computing quantum mechanical average values 
the precise operational sense in which equivalence with a classical 
stochastic process is obtained. We write in the following $y_0,..$ for 
variables and operators in the Schr\"odinger picture. We assume that the wave 
function $\Psi(y_0,~y_0^*,~\eta)$ in the 
Schr\"odinger amplitude (position) representation satisfies (\ref{wkb}).
Namely, we assume a WKB behaviour of $\Psi$ with: 
\begin{equation}
{\hat p}_o~\Psi(y_0,~\eta) = p_{cl}(y_0)~\Psi(y_0,~\eta)~\label{wkb1},
\end{equation}
where $p_{cl}(y)$ is the classical momentum in the large squeezing limit. 
This implies in particular that the probability density 
$|\Psi|^2$ moves along classical trajectories in amplitude space:
\begin{equation}
\frac{\partial |\Psi|^2}{\partial t}= -\sum_{i=1}^2 \frac{\partial}
{\partial y_{0i}}\left(y'_{i,cl}~|\Psi|^2 \right)~.
\end{equation}
Combination of (\ref{rot}) and (\ref{wkb}) then implies 
\begin{eqnarray}
{\cal P}(y_0, y_0^*;~\eta_0) &\equiv & |\Psi(y_0,y_0^*,\eta_0)|^2\nonumber\\
&=& \beta^2(\eta)~|\Psi(\beta y_0,\beta y_0^*,~\eta)|^2\label{P}~,
\end{eqnarray}
where $\beta(\eta)\equiv \frac{|f_k(\eta)|}{|f_k(\eta_0)|}$.

Consider an arbitrary operator $K(y_0,~y^{\dagger}_0)$ in the Schr\"odinger 
representation. Then the quantum expectation value 
$\langle~K(\eta)~ \rangle_q$ at time $\eta$, in the limit where conditions 
(\ref{rot}) and (\ref{wkb}) hold, is given by
\begin{eqnarray}
\langle~K(\eta)~ \rangle_q &=& \int ~dy_{01}~dy_{02} ~K(y_0,~y^*_0)~ 
|\Psi(y_0,~y_0^*,~\eta)|^2\nonumber\\
&=& \int ~dy_{01}~dy_{02}~K(y(\eta),~y^*(\eta))~\beta^2(\eta)~
|\Psi(\beta y_0,~\beta y_0^*,~\eta)|^2\nonumber\\  
&=& \int ~dy_{01}~dy_{02}~K(y(\eta),~y^*(\eta))~
{\cal P}(y_0,y_0^*;~\eta_0)\label{p1}\\  
&=& \int ~de_{y_{1}}~de_{y_{2}}~K(f_k e_y, f_ke_y^*)
 {} ~{\cal W}(e_y,e_y^*;~\eta_0)
\nonumber\\  
&=& \langle~K~\rangle_{cl}~ \ ,
\end{eqnarray}
where ${\cal W}(e_y,~e_y^*;~\eta_0) \equiv f_k^2(\eta_0)~|
\Psi(y_0,y_0^*~\eta_0)|^2$ and $e_y=\sqrt{2k}y_{\bf k}(\eta_0)$ \cite{PS2}.
We have analogously for an operator 
$G(p_0,p_0^{\dagger})\equiv \sum_{n=0}^{\infty} g_{nm}~p_0^n~p_0^{\dagger m}$:
\begin{eqnarray}
\langle~G(\eta)~ \rangle_q &=&
 \int~dy_{01}~dy_{02} ~\Psi^*(y_0,~y_0^*,~\eta)~G(p_0,~p_0^*) 
\Psi(y_0,y_0^*,~\eta)\nonumber\\ 
&=& \int ~dy_{01}~dy_{02}~G(p_{cl}(y_0), p_{cl}^*(y_0^*)) 
|\Psi(y_0,~y_0^*,~\eta)|^2\label{p2}\\
&=& \int ~dy_{01}~dy_{02}~G(p_{cl}(\eta),p_{cl}^*(\eta))
{\cal P}(y_0,~y_0^*,~\eta_0)\label{p3}\\ 
&=& \int ~de_{y_{1}}~de_{y_{2}} ~G(g_{k2}~e_y,g_{k2}e_y^*)
{\cal W}(e_y,~e_y^*;~\eta_0)
\nonumber\\  
&=& \langle~G~\rangle_{cl}.
\end{eqnarray} 
In order to get Eqs. (\ref{p1},\ref{p3}), resp. (\ref{p2}), conditions 
(\ref{P}), resp. (\ref{wkb}), have been used. 

For example, for a two-modes squeezed $N$-particle state, a state containing 
initially $N$ particles for both ${\bf k}$ and ${\bf -k}$, we have
\begin{equation}
{\cal W}(e_y,~e_y^*;~\eta_0)={\cal W}(|e_y|;~\eta_0)=
\frac{1}{\pi}~L_N^2(|e_y|^2)~e^{-|e_y|^2}~.\label{gau}
\end{equation}
In the long wavelength regime $k\ll aH$, one has $|F(k)|\gg 1$
 and we get a WKB state with
\begin{equation}
|F(k)|\gg \frac{1}{2N+1}~.\label{imB}
\end{equation}

Again we see that the system is indistinguishable from a classical stochastic 
process, i.e. we have a probability density ${\cal P}(y_0,~\eta)$ 
whose initial value is fixed by the initial quantum state and which 
evolves in classical zero-measure regions of trajectories in phase space. 
This is really what makes the system so 
peculiar in the limit $|r_k|\to\infty$~: the combination of the WKB behaviour 
as expressed by (\ref{wkb},\ref{wkb1}) {\it together} with the vanishing of 
the decaying mode. As a result, the probability density $|\Psi|^2$ becomes in 
this limit a classical probability density, see 
(\ref{P}) which guarantees the conservation of probability in amplitude 
space for classical paths in the limit of infinite squeezing. It is a crucial 
point that for the equivalent classical stochastic process one has 
regions of trajectories with measure zero in phase space.
Hence, only a probability density in 
{\it amplitude} space is needed and not a joint probability density in 
phase space as already pointed out earlier. 
Were this not the case, quantum coherence of the quantised system would 
resurface at this stage and the latter could not be
indistinguishable from a classical stochastic process.  
\par
We can also look at the problem in a different way which nicely exhibits the 
role of the decaying mode.
For simplicity of notation we shall write henceforth: $dy_0\equiv 
dy_{01}dy_{02},~f(y_0)\equiv f(y_{01},~y_{02})\equiv f(y_0,~y_0^*)$, and 
take a finite volume; $f_0$ denotes the 
quantity $f$ taken at time $\eta_0$. Let us consider the density matrix $\rho$ 
of our state $|\Psi \rangle$. It can be written in the Heisenberg 
representation in the following way:
\begin{equation}
|\Psi,~\eta_0 \rangle \langle \Psi,~\eta_0|= \int \int
 dy_0'~dy_0~\Psi^*(y_0',\eta_0)\Psi(y_0,\eta_0)\vert y_0\rangle\langle
y_0'\vert\ .
\end{equation}
In the Heisenberg representation, the above density matrix will be time 
independent. 
The action of the operator $y({\bf k},~\eta)$ on $|y_0 \rangle$ 
is given by
\begin{equation}
y~|y_0 \rangle = y_{cl}(y_0)~|y_0 \rangle
-\sqrt{\frac{2}{k}}~f_{k2}~p_0~|y_0 \rangle~,   
\end{equation}
and analogously for the operator $p({\bf k},~\eta)$:
\begin{equation}
p~|y_0 \rangle =\sqrt{2k}~g_{k2}~y_0~|y_0 \rangle 
+\sqrt{\frac{2}{k}}~g_{k1}~p_0~|y_0 \rangle~.  
\end{equation}
In the large squeezing regime the second term of both equations becomes  
vanishingly small. When we neglect the decaying mode completely, we get 
\begin{eqnarray}
\langle K(y) \rangle  &=& \int dy_0~K[y_{cl}(y_0)]~|\Psi(y_0)|^2\ ,\\ 
\langle G(p) \rangle  &=& \int dy_0~G[p_{cl}(y_0)]~|\Psi(y_0)|^2\ ,\\
\langle M(y,p) \rangle &=& \int dy_0~M[y_{cl}(y_0), p_{cl}(y_0)]~
|\Psi(y_0)|^2 \ .
\end{eqnarray}
\vskip 2mm

Property (\ref{W2}) of the Wigner function lends itself to the same 
interpretation in the high squeezing limit. 
All expectation values of operators are equal to classical 
averages in phase space, with a distribution $\rho_{cl}(y,p)$ that is equal 
to the Wigner function (\ref{W2}). 
For example, the quantum expectation value of $pp^{\dagger}$ is
\bea
\langle\Psi_{{\bf k}0}|pp^{\dagger}|\Psi_{{\bf k}0}\rangle & = &
2\langle\Psi_{{\bf k}0}|p_1^2|\Psi_{{\bf k}0}\rangle
= 2\langle\Psi_{{\bf k}0}|p_2^2|\Psi_{{\bf k}0}\rangle\nonumber\\
 & = & \frac{1+4F(k)^2}{4|f_k|^2}\ , 
\eea
while the corresponding classical average in the high squeezing limit is
\be
\int\ \D y\D p \ W(y,p)~pp^{\dagger}= \frac{F(k)^2}{\vert f_k\vert^2} \ .
\ee
As in this limit one has $|F(k)|\gg 1$, the difference
between these two expressions becomes negligible.
Thus, even if one could measure operators that involve the
momenta (for example, particle number), and such operators are already 
difficult to measure in laboratory situations \cite{Al}, it would not be 
possible to distinguish the above state from a corresponding classical
stochastic process. The difference could only be observed by measuring 
with extreme precision an operator like $p-p_{cl}$.
Finally we emphasize again that all the results mentioned in this section 
are obtained as a result of the dynamics of the system only.
We therefore have to investigate the sensitivity of our state to the 
environment. Before embarking on this, we shall first consider a simple 
physical system, the free nonrelativistic particle which, surprisingly 
enough, exhibits many analogies with long wavelength fluctuations.
\section{An analogy: The free particle}

A simple, but very illustrative analogy to the cosmological model
presented in Section~2 is the free evolution of a
nonrelativistic quantum particle. Let us first have a look into
the Heisenberg picture. The solution is there
\bea \hat{x}(t) &=& \hat{x_0}+ \frac{\hat{p_0}t}{m}\ , \\
      \hat{p}(t) &=& \hat{p_0} \ . \eea
One recognises that, in the limit of large $t$, position and
momentum approximately {\em commute}, in analogy to (\ref{com})
above. Moreover, in this limit, position is a quantum nondemolition
variable, satisfying the analogous relation to (\ref{QND}).
The crucial difference to the cosmological case is that for the free
particle, the initial {\em position} becomes irrelevant at large
times, whereas for cosmology, it is the initial {\em momentum},
see (\ref{yk}) and (\ref{pk}).

Even more illustrative is the situation in the Schr\"odinger picture.
We start with a Gaussian wave function at time $t_0=0$
of width $b_0$ which has
initial momentum $p_0$ and is centred at $x=x_0$ (we take thus a more
general Gaussian state than the vacuum state of Section 2).
The solution for $t>0$ then reads
\be \psi(x,t)=N(t)\exp\left\{-\frac{\Omega(t)}{2}
     \left(x-x_0-\frac{p_0t}{m}\right)^2+\I \frac{S_0}{\hbar}\right\}\ , \ee
where
\be S_0\equiv p_0\left[x-\frac{1}{2}\left(x_0+\frac{p_0t}{m}\right)\right]
    \lb{S0}   \ee
is a solution to the Hamilton-Jacobi equation. (We have re-inserted
$\hbar$ in all expressions for illustration.)
The explicit expressions for the real and imaginary part of
$\Omega$ read
\bea \Omega_R^{-1} &\equiv& |f|^2 =2b_0^2\left(1+\frac{\hbar^2t^2}
      {4m^2b_0^4}\right)\ , \\
   \frac{\Omega_I}{\Omega_R} &\equiv & -2F=-\frac{\hbar t}{2mb_0^2} \ . \eea
We have introduced the abbreviations $F$ and $|f|$ to facilitate
comparison with (\ref{Psi}). One recognises that large $t$
corresponds to the large $F$ limit, in analogy to the
cosmological case. One also recognises that large $t$ corresponds
to ``large" $\hbar$, so the presence of $\Omega_I$ is a
higher order WKB effect, as is well known. Surprisingly, in the
limit $t\to\infty$, the WKB approximation becomes again valid,
as we shall show now.

Decomposing the wave function into amplitude and phase,
\be \psi(x,t)= R\E^{\I S/\hbar}\ , \ee
one has
\bea S(x,t) &=& S_0 +\frac{\hbar}{2}\arctan(-2F)
   +\frac{\hbar^2t(x-x_0-p_0t/m)^2}{8mb_0^4(1+4F^2)} \\
  & \stackrel{t\to\infty}{\longrightarrow}&
    \frac{mx^2}{2t} + {\cal O}\left(\frac{1}{t^2}\right)\ . \eea
Note that the first term in the second line is independent of
$\hbar$ and an exact solution of the Hamilton-Jacobi equation.
(This solution is found from (\ref{S0}) by construction
of the envelope.)
This is why an exact WKB situation is obtained in this limit,
just as it is obtained for the state (\ref{Psi}). The classical
relation $p=\partial S/\partial x$ immediately yields
$x(t)=p_0t/m$, exhibiting the neglection of $x_0$ in this limit.

The Wigner function reads
\be W(x,p,\eta)= \frac{1}{\pi\hbar}\exp\left(-\frac{(x-x_0-
    p_0t/m)^2}
    {\vert f\vert^2}\right)\ \exp\left(-\frac{\vert f\vert^2}{\hbar^2}
     [p-2F\vert f\vert^{-2}x]^2\right) \ .  \ee 
It is obvious that this is equivalent to (\ref{W1})!
The basic contribution from $x$ and $p$ comes from the elliptical
region in phase space where the negative exponent of the Wigner
function is smaller or equal than one. In the limit of large $t$,
this corresponds to an ellipse that becomes extremely stretched
and tilted (see Figure~1), in full analogy to the cosmological case of long 
wavelength perturbations. In this sense, the free particle exhibits high 
squeezing.

\begin{figure}[htbp]
\epsfxsize=12cm
$$
\epsfbox{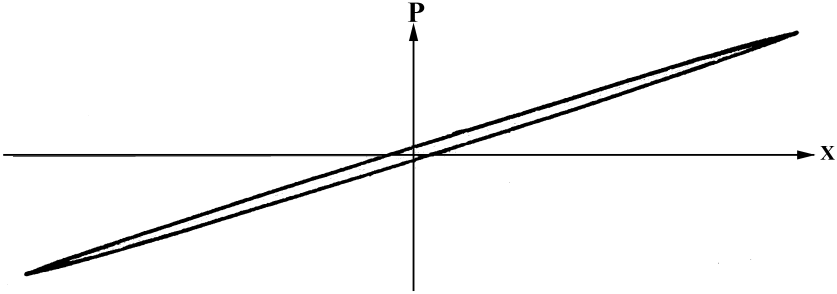}
$$
\caption[]{
The ellipse represents the typical volume in phase space of the free particle 
for large times $t$. The length of the ellipse grows to infinity while the 
width tends to zero. This is equivalent to a classical stochastic process 
with stochastic positions {\it and fixed} corresponding classical momenta. 
For large times $t$, the ellipse tends to the horizontal position.}
\label{figure1}
\end{figure}

Although one has an exact WKB situation for large times, the
corresponding wave function possesses large quantum features in the
sense, that it is very broad in position. In a slit 
experiment, for example, one would expect to obtain notable
interference fringes. However, for $t\to\infty$, the de~Broglie
wavelength goes to infinity, and one would have to increase the size of the 
slit in correspondence to conserve the interference pattern. Remember that 
while the slit is in position, the squeezing is in momentum. 
In order to gain more insight into the pattern on the screen behind the slit, 
we consider now some concrete classical stochastic process. We first specify 
the probability densities ${\cal P}(p_0)$ in the initial momenta and 
${\cal P}(x_0)$ in the initial positions of a free classical stochastic 
particle. Let us consider the following case: 
\begin{eqnarray}
{\cal P}(p_0) &=& \frac{1}{\sqrt{2\pi}~\sigma}~e^{-\frac{p_0^2}
{2 \sigma^2}}\\ 
{\cal P}(x_0) &=& L^{-1}~~{\rm for}~~x_1 - \frac{L}{2}\leq x_0\leq x_1 + 
\frac{L}{2}\ .
\end{eqnarray}
We ask now for the probability ${\cal P}(x(t)\in \Delta x)$ that the free 
particle will have its position at time $t$ in some interval $\Delta x$ 
centered around some arbitrary position $x_1$. 
This probability corresponds to all the particles with initial coordinates 
$(x_0,p_0)$ in phase space inside a tilted area (see Figure~2).

\begin{figure}[htbp]
\epsfxsize=12cm
$$
\epsfbox{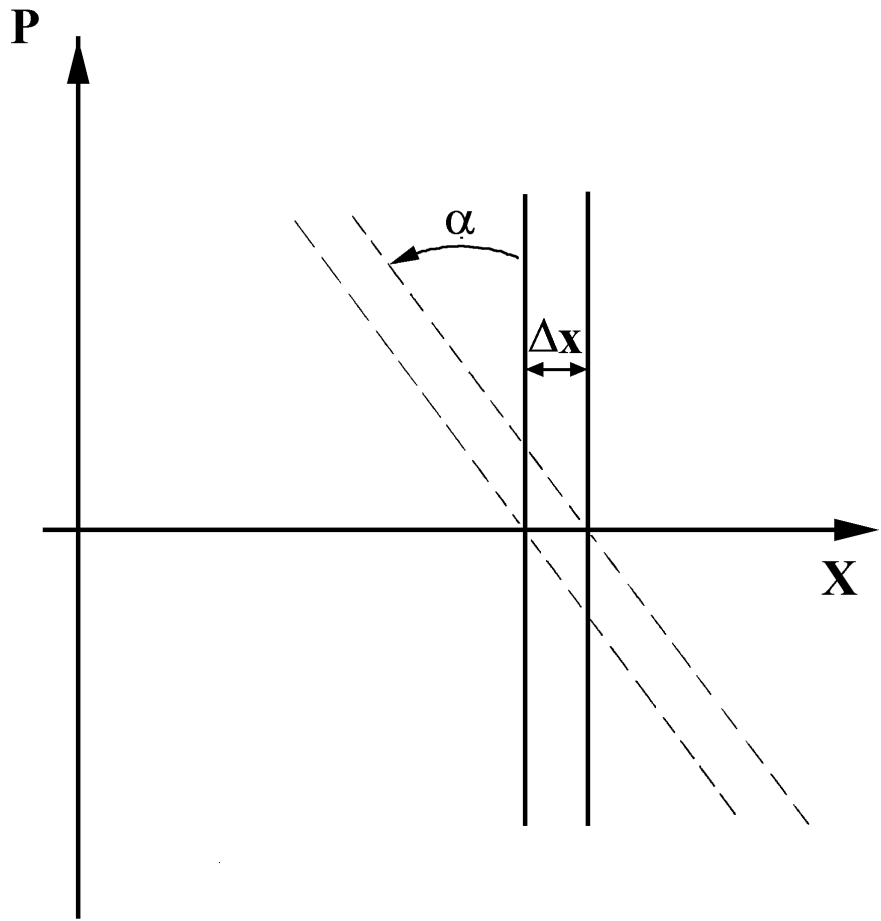}
$$
\caption[]{
The tilted area corresponds to the points in phase space at time $t=0$ that 
will eventually have their position $x(t)$ at time $t$ in some interval 
$\Delta x$. In particular $\alpha = \tan^{-1} \frac{t}{m}$.}
\label{figure2}
\end{figure}

As larger times are considered, this area gets increasingly tilted with 
$\tan \alpha ={t}/{m}$. 
Let us then vary the initial conditions in the positions in such a way that 
$\frac{L}{2}\frac{m}{t}=N \sigma \equiv \Delta p$, with $N$ sufficiently 
large. 
In this way, all the initial momenta 
$-N\sigma\leq p_0\leq N\sigma $ are included in the area and 
if we include even more momenta, the resulting change in 
${\cal P}(x(t)\in \Delta x)$ will be negligible. From the geometry of the 
problem it is now easy to get the result:  
\begin{equation}
{\cal P}(x(t)\in \Delta x)= A \frac{m}{t}~\Delta x~\label{Px},
\end{equation}
where 
\begin{equation}
A = \frac{1}{\sqrt{2\pi}~\sigma}\int\limits^{\Delta p}
_{-\Delta p} e^{-\frac{p^2}{2 \sigma^2}}\approx 1~. 
\end{equation}
The range $\Delta x$ can be taken on the screen; actually, the existence of a 
slit is not important at all for our one-dimensional classical stochastic 
particle. Indeed, all the allowed 
classical trajectories (from different tilted areas) will pass through the 
slit, however at different times. 
Only the width $\Delta x$ is important and not its location on the screen.
The probability per interval $\Delta x$ is constant for given time,
while ${\cal P}(x(t)\in \Delta x)\to 0$ for $t\to \infty$. This 
result is essentially independent from ${\cal P}(x_0)$ because as 
$t \to \infty$ more and more possible initial positions are included 
corresponding to the same momentum $p_0$, and hence (\ref{Px}) is recovered 
independently of the details of ${\cal P}(x_0)$. 
This also corresponds to the fact that as 
$t\to \infty$ the initial positions become negligible compared to 
${p_0t}/{m}$ so that the details of their probability distribution 
becomes irrelevant. We have emphasised in \cite{KPS} that in this limit
the classical arrival time of the free particle is much bigger than
its quantum indeterminacy (which is inversely proportional to the
initial kinetic energy of the particle \cite{AOPRZ}). 
 The result is finally independent of ${\cal P}(p_0)$, 
provided all possible momenta are taken into account.

The quantity ${\cal P}(x(t)\in \Delta x)$ gives us no information about 
${\cal P}(p_0)$; for this we would need a ``momentum'' slit and ask for the 
probability that $p(t)\in \Delta p$ at time $t$, which would just give us 
the probability for $p_0\in \Delta p$. This is in sharp contrast with the 
cosmological perturbations produced during inflation where the various 
observations of quantities like the matter power spectrum $P(k)$ or the 
multipoles $C_{\ell}$ of the cosmic microwave background anisotropies give us 
information about the distribution of the initial amplitudes, the only 
information we have access to. The relevant initial probability distribution 
is in the elongated direction (the amplitudes) for the cosmological 
perturbations, while it is in the squeezed direction (the momenta) for the 
free particle.   

One might wonder whether there are other examples from quantum field
theory in external backgrounds where a similar situation occurs as the case 
of cosmology seems to be very peculiar.
Consider, for example, the case of charged scalar particles in an external
electromagnetic field (a detailed discussion of the functional
Schr\"odinger picture, similar to the framework employed here,
can be found in
\cite{K92}). The analogous equation to (\ref{KG})
reads there
\be \ddot{y}+[m^2+(e{\bf A}+{\bf p})^2]y=0 \  , \ee
where ${\bf A}$ denotes the external vector potential, and $m$
is the mass of the charged (scalar) particle.
One can have parametric resonance in this case too \cite{KLS} when $\bf A$ 
is oscillating. Mathematically, the solution is then the product of an 
oscillating (periodic) part with a growing or decaying exponential function. 
When the decaying solution is neglected, a transition to classical behaviour 
is also obtained. 
%
%
\vskip 2mm

\section{The role of environmental decoherence}

The standard way to understand the emergence of classical
behaviour in quantum theory is, in contrast to the above example,
the interaction of a quantum system with its natural environment
\cite{dec}. The important point in this context is that most quantum objects
cannot be considered as being isolated, but are strongly 
entangled with the states of the environment. The coupling to the
environment may be very small to achieve classical behaviour
as the example of the dust grain interacting only with the
microwave background \cite{dec,JZ} demonstrates. Thus, decoherence
is an ubiquitous phenomenon. 

The actual amount of decoherence depends on the given states and
the interaction. Because of the universal interaction of gravity
with all other fields, the gravitational field
becomes -- apart from small fluctuations, the gravitational waves --
the most classical quantity. This was in particular shown for
the scale factor of a Friedmann Universe whose classicality
is a prerequisite for the classicality of other fields \cite{Ki}.
One may call this the {\em hierarchy of classicality}, with the
(global) gravitational field at the top and small molecules and atoms
at the bottom. This quasiclassical nature of the scale factor
is a necessary condition for the above cosmological scenario,
where the a priori existence of a background spacetime is assumed.

What are the main differences between environment-induced decoherence and 
the above discussed classicality of the field modes?
The squeezed state found above, though it lends itself to a description in 
classical stochastic terms, is nevertheless a quantum state
par excellence (it has, for example, a non-positive Glauder-Sudarshan
distribution function). 
In principle, for large but finite squeezing $|r|$, the coherences are still 
present in the system {\it itself} and could be observed with an appropriate 
experimental setting, though in the case of inflation $|r|$ is large enough 
so that these coherences are unobservable in practice. 
Hence, certainly for the given observational resolution (which eventually 
are those of the satellite missions mentioned in the introduction) but even 
well beyond it, the observations cannot be distinguished from those generated 
by a classical stochastic process as discussed in section 3. 
On the other hand, for a decohered system one is unable
from any practical point of view to observe any 
coherences in the system {\it itself}, although a theoretical, extremely tiny,
coherence width always remains. The reason for this is that the coherences 
are present in the correlations with the environment, and 
the huge number of degrees of freedom in the environment cannot be controlled.

How can the effect of the environment be quantitatively estimated?
A convenient measure for the emergence of quantum entanglement
is the rate of de-separation, first introduced in \cite{KZ}
(see also \cite{dec}). This is defined as follows.
Consider the total system consisting of the relevant part
(our ``system") and irrelevant environmental degrees of freedom.
The total state $|\Psi\rangle$ can be uniquely expanded as a
single sum into (time-dependent) orthonormal basis states of
relevant system and environment -- the Schmidt expansion:
\be |\Psi(t)\rangle =\sum_n \sqrt{p_n(t)}
   |\varphi_n^S(t)\rangle |\phi_n^S(t)\rangle\ , \lb{sch} \ee
where $\{\varphi_n^S\}$ and $\{\phi_n^S\}$ are the Schmidt basis states
of relevant system and environment, respectively.
If at some artificial initial time $t=0$ the total state factorises,
\be |\Psi(0)\rangle =|\varphi_0\rangle|\phi_0\rangle\ , \lb {fac} \ee
the interaction (defined through the Hamiltonian $H$)
will in general lead to entanglement: Up to order $t^2$ one obtains
\bea |\Psi(t)\rangle &\approx& (1-t^2{\cal A})^{1/2}
     |\varphi_0^S(t)\rangle|\phi_0^S(t)\rangle \nonumber\\
   & & \ + (t^2{\cal A})^{1/2}|\varphi_1^S(t)\rangle|\phi_1^S(t)\rangle\ ,
    \eea
where
\be {\cal A} =\sum_{j\neq0,m\neq0}
   \left|\langle\varphi_j\phi_m|H|\varphi_0\phi_0\rangle\right|^2 \ee
is the rate of de-separation ($\{\varphi_j\}$ and $\{\phi_m\}$
denote here a {\em fixed} orthonormal basis for system and
environment, respectively).
It is obvious that for ${\cal A}\neq0$ the total state no longer
factorises. The associated decoherence timescale is thus
$t_D=1/\sqrt{{\cal A}}$. The rate of de-separation can also be related
to a different measure of decoherence -- the increase of local
entropy \cite{dec}. 

If the total Hamiltonain $H$ is of the form
\be H=H_{\varphi}+H_{\phi}+W_{\varphi}\otimes W_{\phi}\ , \ee
${\cal A}$ is independent of the ``free" parts $H_{\varphi}$ and
$H_{\phi}$ and can be written in the form
\be {\cal A}={\cal A}_{\varphi}{\cal A}_{\phi}\ , \ee
where
\bea
{\cal A}_{\varphi}&=&\langle\varphi_0|W_{\varphi}^2|\varphi_0\rangle
-\langle\varphi_0|W_{\varphi}|\varphi_0\rangle^2\equiv
(\Delta W_{\varphi})^2\ , \\
{\cal A}_{\phi}&=&\langle\phi_0|W_{\phi}^2|\phi_0\rangle
-\langle\phi_0|W_{\phi}|\phi_0\rangle^2\equiv
(\Delta W_{\phi})^2\ . \eea
This is just the mean square deviation of the interaction Hamiltonian
$H_{int}\equiv W_{\varphi}\otimes W_{\phi}$
with respect to the initial state. It is this form that we shall use below
for our concrete calculations.

What are the relevant interaction Hamiltonians for fields in the
early Universe? The details are certainly complicated and depend
on the (as yet unkown) precise form of grand unification theories.
What is clear from present particle physics models, however,
is the fact that the interaction is local in {\em field space}
(as opposed to field momentum space).\footnote{Interactions
involving field momenta come from the gravitational part
$G_{abcd}p^{ab}p^{cd}$ in the Hamiltonian constraint and describe
graviton scattering. This is, however, negligible under the
circumstances considered here \cite{KPS}.}
Therefore, field amplitudes are ``measured" by the environment
and one expects large decoherence for states that are broad in
field amplitude space (as happens for the squeezed states
considered here). This is well known from quantum optics
\cite{dec} and will be shown to happen also here. 

To find a lower bound on the amount of decoherence, it is sufficient
to consider a simplified interaction term. We take
\be H_{int}=gk^2(y_{\bf k}z^{\dagger}_{\bf k}+
   y_{\bf k}^{\dagger}z_{\bf k}) =
  2gk^2(y_1z_1+y_2z_2)\ , \ee
where $g$ is a dimensionless coupling constant, and
$z_{\bf k}\equiv z_1+\I z_2$ is the environmental field which is assumed
to possess the same ``free" part as $y_{\bf k}$.
This is similar to the interaction term used in the toy model of
\cite{BLM}.\footnote{In \cite{KPS} we have chosen a slightly different
form, because we did not consider there the complex nature of the
fields.}
 
We assume that at some instant the total state is, as in (\ref{fac}),
a product state of the $y$-part and the $z$-part,
\be \Psi\equiv\psi_{y_1}\psi_{y_2}\psi_{z_1}\psi_{z_2}\ . \lb{yz} \ee
 For the $y$-part
we take our squeezed state produced by inflation, while for the
$z$-part we take for simplicity the vacuum state, i.e., with $\Omega$
in the Gaussian given by $\Omega=\Omega_R=k$ (the essence of the 
result remains unchanged by taking more complicated states).
The rate of de-separation then becomes
\be {\cal A}= (\Delta W_{y_1})^2(\Delta W_{z_1})^2
+(\Delta W_{y_2})^2(\Delta W_{z_2})^2\ , \lb{des} \ee
where $W_{y_1}=\sqrt{2g}ky_1$ etc., and the variances are evaluated
with respect to the various wave functions in (\ref{yz}).
Since 
\[ \langle\psi_{y_1}|y_1^2|\psi_{y_1}\rangle =\frac{|f_k|^2}{2},\;
 \langle\psi_{z_1}|y_1^2|\psi_{z_1}\rangle =\frac{1}{4k}\ , \ldots\ , \]
the rate of de-separation (\ref{des}) is given by
\be {\cal A}=g^2k^3|f_k|^2= \frac{g^2k^2}{2}(\cosh 2r_k+
   \cos 2\varphi_k\sinh 2r_k)\ . \ee
The measure for quantum entanglement is thus essentially given by the power 
spectrum of the fluctuations!
In the limit of large squeezing in $p$-direction ($\varphi_k\to 0$)
this becomes
\be {\cal A}\to \frac{g^2k^2}{2}\E^{2r_k}\ . \ee
The corresponding decoherence time is then given by
\be t_D\sim \frac{a}{gk\E^r} \sim \frac{\lambda_{phys}}{g\E^r}\ ,
       \lb {dt} \ee
where $\lambda_{phys}$ denotes the physical wavelength of the
fluctuations.\footnote{The factor $a$ in (\ref{dt}) occurs after
the physical fields $\phi=a^{-1}y$ etc. are considered.}
For example, for $\lambda_{phys}\sim 10^{28}\mbox{cm}$
(present horizon scale) and $\E^r\sim 10^{50}$ (squeezing factor
of this mode) one has
\[ t_D\sim 10^{-31}g^{-1}\mbox{sec}\ , \]
so that decoherence would be negligible only if one fine-tuned the
coupling to values $g\ll 10^{-31}$ -- a totally unrealistic fine-tuning!

It is straightforward to extend this analysis to more complicated
couplings. Taking, for example, a quadratic coupling,
\[ H_{int}=2gk^3(y_1^2z_1^2+y_2^2z_2^2)\ , \]
one finds in the limit of large squeezing for the rate of de-separation
\be {\cal A}\to \frac{9}{32}g^2k^2\E^{4r} \ee
which is much bigger than the expression for linear coupling.
The decoherence time for the scales which now appear inside the horizon
is then
\be t_D\sim 10^{-81}g^{-1}\mbox{sec}\ , \ee
and one would have to fine-tune $g$ even more to have negligible
decoherence. 
We note that $t_D$ is proportional to the wavelength of the modes
because localisation becomes worse for larger wavelengths
\cite{dec,JZ}. This is, however, largely overcompensated by the
high squeezing of these modes due to inflation (the factor
$\E^{r}$ in the denominator of $t_D$). 

Due to the nature of the interaction, which is local in field
space, the pointer basis defining classical properties is
the field amplitude basis and not, for example, the particle number basis.
This basis is stable during the dynamical evolution
because of the quantum nondemolition condition (16).
An analogous example in quantum electrodynamics
has been discussed in \cite{Ki3,AZ}.

To go beyond estimates, one has to take realistic models and
calculate the corresponding decoherence times quantitatively,
for example by the use of the influence functional method
(an introduction to this method can be found in Chap.~5 of \cite{dec}).
Early investigations of decoherence for primordial fluctuations
are \cite{BLM} and \cite{SK}; more recent and more refined 
calculations can be found in \cite{Hu,BV} and the references therein.
It is, however, of fundamental importance that in order to leave the main 
predictions of inflation unaltered, interaction with the environment should 
{\it not} destroy the fixed phase of the perturbations, though it can, and 
certainly will, affect the coherence between growing and decaying mode. 
In particular, it is crucial that environment-induced decoherence does not 
produce a density matrix which is diagonal in the basis of number eigenstates 
\cite{LPS}, as assumed in some of the above quoted references, because in 
this case the phase becomes random and some basic predictions concerning the 
CMB anisotropies are dramatically changed (see conclusion).

It is also interesting to calculate the rate of de-separation, if our
system is initially in the number eigenstate (26), while for the
environmental ($z$-) part the vacuum state is retained. A straightforward
calculation using the expressions presented in \cite{PS2} yields
\be {\cal A}_N= (2N+1){\cal A}\ , \ee
with ${\cal A}$ given by (78). Therefore, the rate of de-separation is
even stronger than in the vacuum case, as expected.

For Schr\"odinger cat states (such as the states used recently in
quantum optical experiments of decoherence \cite{Ha}) one would expect
an even higher rate of de-separation and even smaller decoherence
timescale (cf. Sect.~3.3.3 in \cite{dec}).

A good analogy to our case of primordial fluctuations is 
Fermi's golden rule and the exponential decay in quantum
mechanics \cite{dec}. For isolated systems, this is known to
hold only approximately, although deviations are hard to measure and
have been observed only recently (this would in our example
correspond to observe effects of the decaying mode).
 If, however, the environment
is taken into account in this case, exponential decay is enforced
by this interaction and no deviations from it can be seen.
At the same time, all probabilities remain unchanged.
(This would correspond to the impossibility to observe any coherence
effects for the primordial fluctuations, while at the same time
all probabilities remain unchanged).

We want to finally emphasise again that decoherence can be of importance
(due to the emergence of quantum entanglement) long before any dynamical
back reaction occurs. This is the reason why all the predictions
concerning the primordial fluctuations are unchanged, while
at the same time nonclassical interference terms remain essentially suppressed 
in the system itself. 
Quantum mechanical examples tell that ``decohered" wave packets show no 
interference even if they occupy
the same region of space (see in particular Fig.~3.7
in \cite{dec}). Although nonclassical behaviour of
squeezed states \cite{SW} is difficult to observe for an isolated
system (and $|r|$ much smaller than for inflation, otherwise it is hopeless), 
it is in practice impossible to observe for the decohered system. 

\vskip 2mm

\section{Conclusions}

We have investigated in detail the quantum-to-classical transition of the 
fluctuations of quantum origin produced during inflation. When no 
interaction with the environment is taken into account, such a transition 
takes place up to a precision well beyond observational capabilities. This 
is directly related to the fact that it is
 possible to describe the fluctuations 
nowadays solely with the help of the ``growing'' quasi-isotropic mode.  
This transition means that the quantum coherence can be expressed in 
classical terms, namely the system can be described as a {\it stochastic} 
classical system.
That this is very far away from a classical (deterministic) system is very 
clear in the example of a free particle: at very large times, one cannot 
ascribe anymore to it a definite trajectory in phase space, but 
rather one has a {\it classical} probability density with stochastic 
amplitudes (positions) $x$ and fixed momenta $p_{cl}(x)$.
The initial quantum state then completely defines the statistics of the 
fluctuations through the probability distribution $|\Psi|^2$. Most 
inflationary models lead to a Gaussian statistics of the fluctuations, a 
result in good agreement with observations. 
Clearly this is very far away from a classical free particle! This aspect is 
somehow hidden in the case of cosmological fluctuations because in the latter 
case one is willing to accept the stochasticity of the fluctuations,
and it would 
look absurd to even try a deterministic description of these fluctuations, 
even if one believes the fluctuations are classical from the very beginning. 
This explains why the description in terms of a classical stochastic 
process does not look surprising. It is only when one thinks of the 
quantum origin of the fluctuations that the peculiar quantum nature appears.   
We stress also that the quantum-to-classical transition is a result of the 
expansion of the universe and that it depends on the stretching of the 
fluctuations while they are outside the Hubble radius.
Note that this would not apply to scalar fields with too large mass~\cite{Mi}, 
in complete accordance with the fact that these fields cannot be described by 
just a growing mode. 

We again emphasise that the environment has to be taken into account,
since the highly squeezed states are extremely sensitive to the
presence of an environment, as has been discussed in Sect.~5. 
When it is taken into account, even coherences which are unobservable in 
practice but still present {\it in} the system, 
essentially disappear from the system itself, since they are 
``hidden'' in the correlations with the huge number of degrees of freedom of 
the environment. However, in the peculiar case of inflation these coherences, 
not expressible in classical stochastic terms, are anyway tremendously tiny. 
It is not even clear that environment-induced decoherence would be effective 
enough to reduce them any further. However, interaction with the environment 
has an irreversible character and is certainly crucial regarding the problem 
of the entropy of the fluctuations.
It is crucial that this interaction does not spoil the standard predictions of 
inflationary physics which will be possibly tested in the near future by 
the satellite missions {\it MAP} and, with even higher accuracy, by 
{\it PLANCK Surveyor}. For example, the fact that the fluctuations 
have stochastic amplitudes but {\it fixed} phases results in the appearance 
of (Sakharov or Doppler or acoustic) peaks on small angular scales in the 
angular power spectrum of the cosmic microwave background anisotropies. 

We note that our discussions exhibit a surprising connection between
cosmology -- the origin of structure -- and fundamentals of quantum
theory \cite{KPS}. The quantum-to-classical transition by decoherence is a 
very general process as studied recently in quantum optical experiments 
\cite{Ha}.

We emphasised in Sect.~2 that the high squeezing of
the quantum state for the primordial fluctuations is ``generic''.
One can, of course, start with any ``quantum state" at the end of inflation
(not necessarily highly squeezed) and evolve it
back to the beginning of inflation by the Schr\"odinger equation, where
it yields an acceptable initial state. However, this state should 
initially (before inflation) be tremendously {\em narrow} in $y$ and may be 
rejected as being unnatural (this is our {\em quantum no hair conjecture}).
Assuming that such an initial state is self consistent with inflation, 
which is certainly {\it not} the case for most models, then the 
longer the inflationary phase, the better our conjecture is 
expected to work; however, the minimum duration required for inflation to be 
of cosmological interest is certainly effective enough in this respect. 
The high squeezing of the fluctuations
and the ensuing quantum-to-classical transition is a generic feature of the 
inflationary phase itself. 

\section*{Acknowledgements}
We thank Alexei Starobinsky for many illuminating discussions.
C.K. acknowledges financial support by the University of Tours
 during his visit to 
Tours, and D.P. acknowledges financial support by the DAAD during
his visit to Freiburg.

\end{document}